\documentclass[journal]{IEEEtran}

\ifCLASSINFOpdf
\else
   \usepackage[dvips]{graphicx}
\fi
\usepackage{url}

\usepackage{graphicx}
\usepackage{amsmath,amssymb,amsfonts}
\usepackage{subfig}
\usepackage{booktabs,multirow}
\usepackage{bm, cite}
\usepackage{algorithm, algpseudocode}
\begin{document}

\title{\centering{Accelerometric Method for Cuffless Continuous Blood Pressure Measurement}}

\author{Mousumi Das, Tilendra Choudhary, L.N. Sharma, and M.K. Bhuyan
\thanks{M. Das, T. Choudhary, L.N. Sharma, and M.K. Bhuyan are with the Department of Electronics and Electrical   Engineering, Indian Institute of Technology Guwahati, India-781039 (E-mails: \{mousumi18a, tilendra, lns, mkb\}@iitg.ac.in).}}

\maketitle

\begin{abstract}
Pulse transit time (PTT) has been widely used for cuffless blood pressure (BP) measurement. But, it requires more than one cardiovascular signals involving more than one sensing device. In this paper, we propose a method for continuous cuffless blood pressure measurement with the help of left ventricular ejection time (LVET). The LVET is estimated using a signal obtained through a micro-electromechanical system (MEMS)-based accelerometric sensor. The sensor acquires a seismocardiogram (SCG) signal at the chest surface, and the LVET information is extracted. Both systolic blood pressure (SBP) and diastolic blood pressure (DBP) are estimated by calibrating the system with the original arterial blood pressure values of the subjects. The proposed method is evaluated using different quantitative measures on the signals collected from ten subjects under the supine position. The performance of the proposed method is also compared with two earlier approaches, where PTT intervals are estimated from electrocardiogram (ECG)-photoplethysmogram (PPG) and SCG-PPG, respectively. The performance results clearly show that the proposed method is comparable with the state-of-the-art methods.
Also, the computed blood pressure is compared with the original one, measured through a CNAP system. It gives the mean errors of the estimated systolic BP and diastolic BP within the range of $-$0.19$\pm$3.3 mmHg and $-$1.29$\pm$2.6 mmHg, respectively. The mean absolute errors for systolic BP and diastolic BP are 3.2 mmHg and 2.6 mmHg, respectively. The accuracy of BPs estimated from the proposed method satisfies the requirements of the IEEE standard of 5$\pm$8 mmHg deviation, and thus, it may be used for ubiquitous long term blood pressure monitoring.
\end{abstract}


\maketitle

\section{Introduction}
\label{sec:introduction}
\IEEEPARstart{H}{igh} blood pressure is a major risk factor that leads to serious cardiovascular diseases (CVDs), such as hypertensive heart disease, stroke, and coronary artery disease \cite{b1}. Therefore, effective and regular monitoring of blood pressure (BP) is necessary for early detection of hypertension. This helps clinical maintenance of the CVDs.  
To monitor the continuous BP in real time in an intensive care unit (ICU), the most common method is to insert a cannula needle in a suitable artery. For arterial blood pressure measurement, it is a gold standard method. Because of its invasive nature, it may introduce some other clinical complications, such as bleeding, infection, and ischemia. Also, it requires a standard clinical setting and expert. Among other modalities, mercury sphygmomanometer is considered as the most accurate non-invasive device. But, it requires an inflatable cuff that may cause discomfort and pain to the user. Moreover, it only provides the instantaneous blood pressure, and thus, cannot be used for long term ambulatory BP monitoring. Arterial tonometry and arterial volume clamp are other primary techniques to obtain continuous BP in non-invasive way, but they need occlusive cuff \cite{b2}. All these devices  have limitations for continuous and long term BP measurement. Therefore, there is a need of advance technology that can provide continuous BP measurement.

\begin{figure*}
\centering
\includegraphics[width=7.25in]{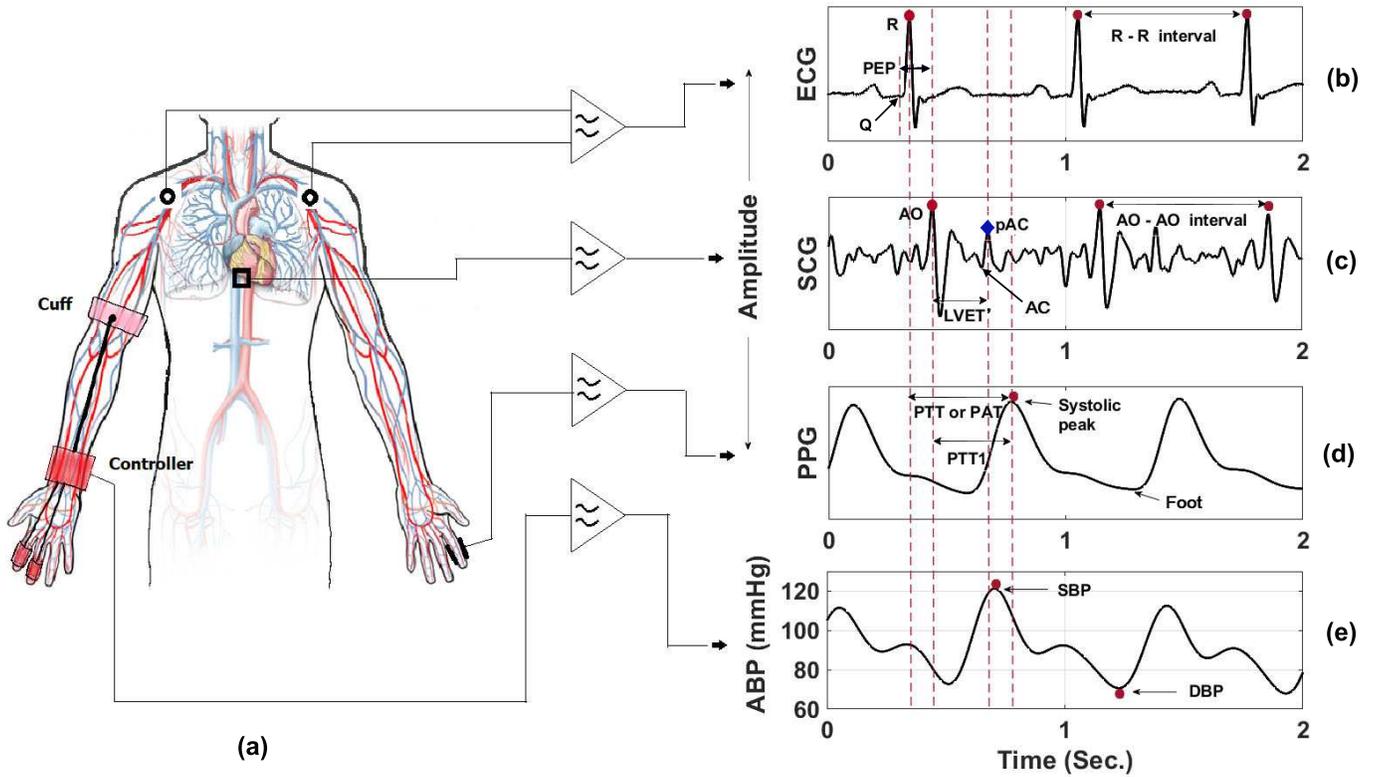}
\caption{Simultaneously recorded cardiological signals and their annotations: (a) Sensor placement on human body, (b) ECG signal, (c) SCG signal, (d) PPG signal, and (e) ABP signal.}
\label{Feature_annotation}
\end{figure*}

In recent years, research on cuffless BP measurement methods for continuous BP monitoring has gained much attention. In literature, it is found that the pulse transit time (PTT) is employed to estimate blood pressure. The PTT is defined as the time taken for a cardiac pulse in artery to travel from a proximal point to a distal point. Usually, for the calculation of PTT interval, R-peak of electrocardiogram~(ECG) and systolic peak or onset of photoplethysmogram~(PPG) are used \cite{b2,b8,b9,b10,b11,b12}. Recently, some of the BP estimation methods have introduced the use of concurrent seismocardiogram (SCG) and PPG signals for the PTT estimation \cite{b5,b14,b15}. For this purpose, they suggested aortic valve opening (AO) peak of the SCG as a proximal point, and systolic peak or onset of the PPG as a distal point. All of these approaches require two correlated cardiac pulse  signals, either ECG-PPG or SCG-PPG. Acquiring more than one signal and estimating the location specific feature points is computationally complex. It also reduces the flexibility of the system due to the involvement of different sensing modalities. This type of complex system involving two or more signals to measure the BP parameters may also be expensive. However, the SCG components of two different orthogonal axes have also been investigated \cite{b13}, where the components are acquired from a single sensor. The method relies on time-interval from systole-profile onset in x-axis to diastole-profile onset in z-axis of SCG for systolic blood pressure (SBP) estimation. However, determination of these onset points is a crucial task. Additionally, achieving a good accuracy is still a problem. The research on BP estimation using SCG signals is still far from maturity and has not yet advanced to a stage where it can be successfully deployed as a medical user interface in consumer-grade applications. The proposed method is  the first ever one of its kind where BP is estimated using a single SCG component. Looking at the above facts, the main aim of this proposed work is to estimate blood pressure using a single signal component acquired through a micro-electromechanical system (MEMS)-based accelerometric device. A tri-axial accelerometer is used to record the SCG signal. However, the proposed method does not rely on SCG components of $x$- and $y$-axis. Only the $z$-axis component is used for processing. Seismocardigraphy is a non-invasive technique used to measure vibrations on the chest wall induced by heartbeats and cardiac movements \cite{b3}. The SCG signal is used to determine the left ventricular ejection time (LVET). Then, it computes the BP metrics using a calibration procedure. Thus, we propose  a low cost, light weight, portable as well as user friendly device for continuous cuffless BP measurement instead of using multi-modal system. Also, recording process of SCG is simpler and provides comfortability to the user.


\section{SCG Signal and its Relation to BP}
The generation of chest-wall vibrations is mainly caused due to change in volume, pressure and shape of the heart during different stages of a cardiac cycle. The chest wall vibrational data can be non-invasively acquired by placing a MEMS-based tiny tri-axial accelerometer sensor on the precordial area of the chest, generally at the lower end of the sternum on the xiphoid process \cite{b3}. The SCG captures mechanical activities of the heart, such as opening and closing of valves, isovolumetric moment, isotonic contraction, rapid blood  filling and its ejection. It is expected that an SCG signal can give more clinically useful information as compared to other cardiac signals. In Fig. \ref{Feature_annotation}, the SCG signal is shown along with other concurrent ECG, PPG and arterial blood pressure (ABP) signals.

During major left ventricular depolarization through purkinje fibers, the mitral valve is closed, and in a standard ECG, the R-wave is seen for this phase (Fig. \ref{Feature_annotation}). After the occurrence of R-peak in the electrocardiogram, further contraction of myocardium increases the pressure in the left ventricle. The first heart sound, `$\mathrm{S1}$', is produced due to closure of the mitral and tricuspid valves. At this moment, aortic valve is not opened. The ventricle works as a closed chamber. The volume remains the same, the pressure increases rapidly due to rapid contraction. This phase is known as isovolumetric contraction. The period until the opening of aortic valve from the depolarization of the left ventricle is known as pre-ejection period (PEP). When the ventricular pressure reaches to the aortic pressure, the aortic valve opens (AO peak in SCG) and blood flows out of the aorta. At this instant, pressure pulse is initiated in aorta for systematic circulation through arteries. To calculate the PTT, the R-peak is considered as the proximal point in the literature. However, in reality, R-peak represents the depolarization of left ventricles, which does not imply the flow of the blood. If the R-peak is considered as proximal point, it includes the PEP interval and may result an incorrect estimation of the PTT. Therefore, it is suggested to consider aortic valve opening point as a proximal point at which the blood starts flowing. Once the ejection of blood from ventricle is completed, the aortic valve is closed. Due to closure of pulmonary and aortic valve, the second heart sound `$\mathrm{S2}$' is produced. But, at this point, the mitral valve is not opened and ventricle is in isovolumetric relaxation phase. Due to re-polarization of ventricles (T-wave in ECG), the ventricular pressure falls down quickly. Because of this, the mitral valve is opened and pressure pulse in the arteries is died out completely at AC point in SCG. During each heart cycle, the blood pressure pulse fluctuates between its  maximum and minimum values. Thus, it is justified to compute the LVET using AO-AC points of the SCG trace. This may resolve the existing anomalies  in various methods for the estimation of PTT.

Similar to the PTT, it is obvious that the LVET is a function of pressure, and so, the pressure parameters can be estimated from it.  Within the time-period between AO and AC, the blood flows throughout the body, and thus, blood pressure pulse reaches peripheral sites. So, the occurrence of pulse at peripheral sites takes place only within the systole period. Thus, the end of the systolic period, \textit{i.e.,} AC can be considered as the distal point. Once LVET is calculated, the BP can be easily estimated through a calibration method.

The paper is organized as follows:  in Section III, we present the proposed cuffless and continuous BP measurement method. Section IV presents the experiments and the results, followed by conclusion in Section V.

\section{Proposed Method}
During the systolic phase of the cardiac cycle, blood is pumped out from the heart. It exerts force on the wall of arteries in systematic circulation and it is measured as blood pressure (BP). In a normal subject, the systolic and diastolic pressure is clinically written as:
\begin{equation} 
\frac{\mathrm{Systolic\,blood\,pressure}}{\mathrm{Diastolic\,blood\, pressure}}=\frac{SBP}{DBP}=\frac{\mathrm{120\, mmHg}}{\mathrm{80\, mmHg}}
\label{equation-1}
\end{equation}
The computation of SBP and DBP is based on the principle of pulse propagation through major arteries. 
The systolic and diastolic blood pressure in a major artery are directly proportional to the total cardiac output ($Q$) and total peripheral resistance ($\Omega$). The $\Omega$ includes resistances of the arteries, arterioles, capillaries, venules, and veins in systematic circulation.  The BP is given as:
\begin{equation} 
BP=Q\times\Omega
\label{equation-2}
\end{equation}
The cardiac output, $Q$, mainly depends on heart rate (HR) and stroke volume (SV). Therefore, it is expressed as:
\begin{equation} 
Q=HR \,\times \, SV
\label{equation-3}
\end{equation}
The $SV$ depends on end systolic volume ($SV_{l}$) and end diastolic volume ($SV_{m}$) of a ventricle with a relation, 
\begin{equation} 
SV=SV_{l}-SV_{m}
\label{equation-4}
\end{equation}
The LVET gives the systolic time intervals that corresponds the interval between ejection and termination of the aortic flow. Thus, the LVET is directly influenced by the SV. The LVET is expressed as:
\begin{equation} 
LVET = AC-AO
\label{equation-5}
\end{equation}
where,  $AO$ and  $AC$ represents aortic valve opening and closing instants, respectively. At the $AO$ instant, the blood is ejected out of the heart, and therefore, can be considered as a proximal point for a pressure pulse. At the $AC$ instant, the blood reaches the measurement sites and returns back from the whole body to the heart, and therefore can be considered as a termination point of the pulse. This justifies the use of LVET in place of PTT to calculate BP. Also, the LVET is inversely related to HR \cite{b18}. It is difficult to identify AC due to large morphological variability of SCGs among the subjects, and the SCG is susceptible to the distortions caused due to body movements, respiration  and other environmental noises \cite{b17}. The proposed model uses the timing-information of pAC peak (peak just after AC point as shown in Fig. \ref{Feature_annotation}) for the computation of LVET interval.  With this modification, an approximation of LVET, say LVET$'$, is used for modelling the BP in the proposed method. The LVET$'$ is computed as:
\begin{equation} 
LVET' = pAC-AO
\label{equation-6}
\end{equation}

\noindent 
By utilising the logarithmic LVET$'$ and HR information, the BP is  modelled as: 
\begin{equation} 
BP = a \cdot ln(LVET') + b \cdot HR +c
\label{equation-7}
\end{equation}
where, $a$, $b$, and $c$ are used to parametrize the proposed model. These parameters are subject-specific and are estimated via modelling in a least square sense.
The $HR$ is calculated from the time interval between consecutive AO peaks. The proposed model can be represented in a vector-matrix form as:
\begin{equation}
\begin{bmatrix}BP_{1} \\ BP_{2} \\ . \\.\\.\\ BP_{n}\end{bmatrix} =
\begin{bmatrix}ln(LVET'_{1}) & HR_{1} & 1\\ ln(LVET'_{2})& HR_{2}& 1\\ . & . &. \\.&.&.\\.&.&.\\ln(LVET'_{n}) & HR_{n}& 1\end{bmatrix}\begin{bmatrix}a \\ b \\ c
\end{bmatrix}
\label{equation-8}
\end{equation}
where, $n$ is the total number of beats, the $BP$ vector with dimension $n\times1$, represents the original arterial blood pressure and it is used for training the model for both SBP and the DBP estimation. The $LVET'$ and $HR$ are the parameters that are estimated simultaneously for each beat along with the original BP and represented by a $n\times3$ matrix . The unknown coefficients $a$, $b$ and $c$ are represented by a $3\times1$ vector. The unknown vector is estimated using linear regression by applying pseudo inverse as:
\begin{equation}
X=(A^{T}.A)^{-1}.A^{T}.BP
\label{equation-9}
\end{equation}
where, $X$ is the unknown coefficient vector and  $A$ is the matrix that carries the information related to the estimated $LVET'$ and $HR$ values. The coefficients are estimated for each individual for both SBP and DBP, separately.
\begin{figure}
\centerline{\includegraphics[width=3.5in]{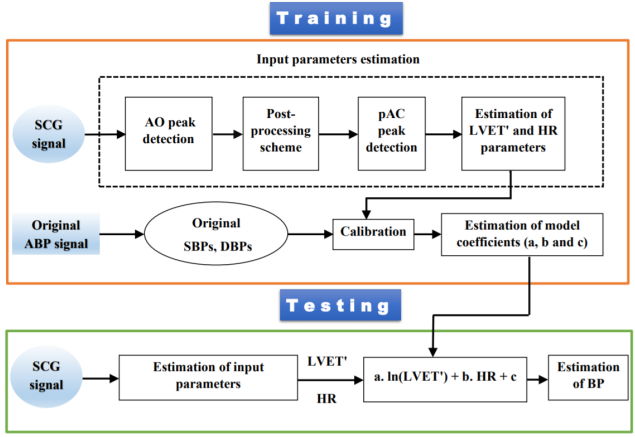}}
\caption{Block diagram of the proposed BP estimation method.}
\label{Proposed System}
\end{figure}
Fig. \ref{Proposed System} shows a block diagram of the proposed cuffless BP estimation algorithm. It has training and testing phases. The training phase is required only during initialization where the coefficients $a$, $b$ and $c$ are computed using the regression method. Then, the testing phase becomes independent of BP parameters for a personalized measurement.
The steps involved in the proposed method are discussed in the following subsections.

\subsection{AO Peak Detection}
Initially, the AO peaks in the SCG are detected using our earlier proposed VMD-based scheme \cite{b16}.  The pseudocode is given in Algorithm \ref{alg:code1}. Although the method provides a robust detection of the AO peaks, it produces misdetections and false-detections in some critical cases. To address this issue, a post-processing scheme is proposed.

\begin{algorithm}[!t]
\footnotesize
\caption{AO instant detection in an SCG signal}
\label{alg:code1}
\renewcommand{\algorithmicrequire}{\textbf{Initialization:}}
\renewcommand{\algorithmicrequire}{\textbf{Input:}}
\renewcommand{\algorithmicensure}{\textbf{Output:}}

\begin{algorithmic}[1]

\Require SCG signal $x[n];~n=1,2,.....,\mathrm{N}$


\hspace{-1.2cm}
\underline{\textbf{Low-Frequency artifacts removal }}

\State Decompose signal $x[n]$ using MVMD 

\Statex ~~$[m_{i}^{I}, \omega_{i}^{I}] = \text{VMD}(x,\alpha_{1},K_{1})$; \qquad  $i = 1,2,...K_{1}$

\Statex  // where  $m_{i}^{I}:=$ decomposed $i^{th}$ mode of $x[n]$, 

\Statex ~~$\omega_{i}^{I}:=$ central frequency of $m_{i}^{I}$, 
 
\Statex ~~$\alpha_{1}:=$~allowing mode-bandwidth, 
 
\Statex ~~$K_{1}:=$~number of decomposed modes 

\State Extract the detrended SCG signal 

\Statex ~~$x_{d}[n]=x[n]-\sum_{i} m_{i}^{I}[n]$ 

\hspace{-1.2cm} 
\underline{\textbf{Systolic Profile Enhancement}}

\State Resolve useful signal components using MVMD
 
\Statex ~~$[m_{j}^{II}, \omega_{j}^{II}]$ = VMD$(x_{d}, \alpha_{2}, K_{2})$; \qquad  $j=1,2,...K_{2}$  

\State Extract Gaussian derivavtive filtered modes (GDFMs) 
 
\Statex ~~$g_{j}[n] = d[m] \circledast m_{j}^{II}[n]$; \qquad $m=1,2,...L-1$

\Statex // where Gaussian derivative kernel $d[m]:= g[m+1] - g[m]$,  
 
\Statex ~~Gaussian window $g[l]:=\text{exp}\left(\frac{-1}{\sigma^{2}}\left[\frac{l}{(L-1)/2}\right]^{2}\right);$  $0\leq\mid l\mid\leq \dfrac{(L-1)}{2} $

\State Compute relative GDFM energy (RGE) for mode selection 
 
\Statex ~~RGE: $e_{j} = \dfrac{E_{g_j}}{\sum_{j} E_{g_j}}; \qquad j= 1,2....K_{2}$
 
\Statex ~~Reconstructed signal: $s[n]= e_{j^*}^{2}g_{j^*} + \text{P} e_{j^*-1}^{2}g_{j^*-1} + \text{Q} e_{j^*+1}^{2}g_{j^*+1} $

\Statex // where $e_{j^*}$ denotes maximum RGE and $j^*$ is corresponding mode index
 
\Statex ~~P=1, Q=0 $\leftarrow$ if $\mid e_{j^*} - e_{j^*-i}\mid < \rho; \qquad \rho \in [0,1]$
 
\Statex ~~P=0, Q=1 $\leftarrow$ if $\mid e_{j^*} - e_{j^*+i}\mid < \rho;$
 
\Statex ~~P=0, Q=0 $\leftarrow$ otherwise 

\hspace{-1.2cm} 
\underline{\textbf{AO peak approximation}}

\State Construct envelope on systolic profiles 
 
\Statex ~~Difference envelope: $D_{env} = U_{env} - L_{env} $ 

\Statex // where $ U_{env}, L_{env}$: upper and lower envelopes of s[n], respectively
 
\Statex ~~Thresholding: $T_{env}[n] = (D_{env}^2[n]> \tau) \times D_{env}[n];\tau>0 $ 

\State  Find zero-crossing locations
 
\Statex ~~Approximated AO peaks: $a[v]\leftarrow$ Positive zero crossings of $\widehat{T}_{env}[n]$

\Statex // where $\widehat{T}_{env}$ corresponds Hilbert transform of $T_{env};$ and $v=1,2,...,V$

\State Estimate True AO peaks using cardiac cycle envelope (CCE)
 
\Statex ~~Construct CCE using absolute operator and triple integration on GDFMs
 
\Statex ~~True AO peaks: $\widetilde{a}[k]\Longleftarrow$ $a[v]$ elements lying within the range $\mathbb{Q}$

\Statex // where $\mathbb{Q} \in (\text{CCE\_peak}_{k}, \text{CCE\_peak}_{k} + 350ms)$; $~k=1,2,.....,\mathrm{K}$

\Ensure Detected AO peak locations $\widetilde{a}[k]$
\end{algorithmic}
\end{algorithm}


\begin{figure}
\vspace{-1cm}
\centering
\includegraphics[width=3in]{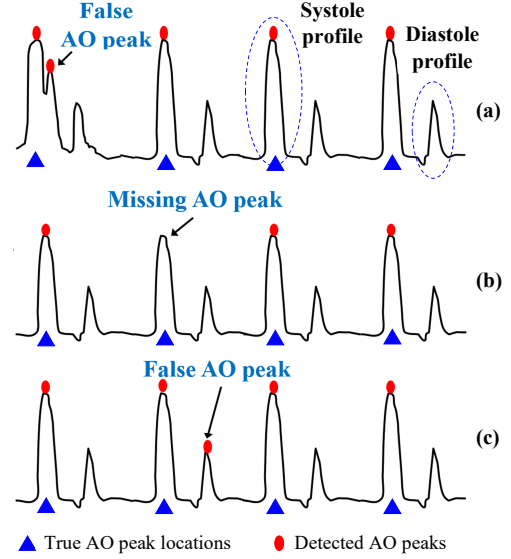}
\vspace{-2cm}
\caption{SCG envelopes showing false alarming conditions in AO peak detection. (a) Case 1: Missing AO peak, (b) case 2: false detection on systole profile, and (c) case 3: false detection on diastole profile}
\label{fig_lossrate}
\end{figure}

\subsection{Post-processing for AO Peak Detection}
A post-processing technique is developed and employed in this algorithm (Algorithm \ref{alg:code1}) to improve the performance of the AO peak detection by reducing the false alarming conditions.
 
Let, $\bm{pks} = [p_{1}, p_{2},  \cdots, p_{n}]$ is a vector of detected AO peaks in a SCG segment.
At first, the difference of each consecutive AO peak instants is calculated as: 
\begin{equation}
d_{i} = p_{i+1}- p_{i}, 
\end{equation}
where, $i$ ($i = 1,2, \cdots, n-1$) denotes the peak index.
Then, median-difference of these AO instants is computed as a reference, say $M$.
The error between $d_{i}$ and $M$ is compared with a threshold (say $TH$). 
Usually, the following three false alarming situations are encountered in the AO peak detection process. These cases are shown through the SCG envelopes in Fig. \ref{fig_lossrate}.
\begin{enumerate}
\item \textbf{Missing AO peaks:}
This false negative condition is identified if , $(d_{i} - M) > 3 \times TH$, and it can be avoided by determining the maxima between $p_{i}+\tau$  and $p_{i+1}-\tau$ in the SCG, where $\tau = 60$ ms.

\item \textbf{False detection on systole profile:}
An AO peak detected in the systole profile is considered as false positive, if $(d_{i}- M) < - 0.7 \times M$, and the condition can be avoided by discarding $p_{i+1}$ peak.

\item \textbf{False detection on diastole profile:} 
After taking care of the first two conditions, the $\bm{pks}$ and $\bm{d_{i}}$ vectors are updated. Subsequently, the method looks for the false positive in the diastole profile by checking the condition, $[(d_{i}- M) < -TH$  and  $(d_{i+1}- M) < -TH]$, and if the condition is satisfied, $p_{i+1}$ peak is removed.
\end{enumerate}

The criterion mentioned above were designed with empirical analysis. In this way, an AO peak correction strategy is adopted in our proposed method. After that, pAC peaks on the SCG signal are detected to measure the LVET$'$ interval.

\subsection{pAC Peak Detection}
The pAC peaks are detected using a technique given in \cite{b17}. At first, it divides each of the consecutive AO-AO intervals at their midpoint ($M_{1}$), which is expressed as:
\begin{equation}
\text{$M_{1}$} = \dfrac{p_{i} + p_{i+1}}{2}\label{equation}
\end{equation}
where, $p_{i}$ and $p_{i+1}$ are consecutive AO peaks from the updated $\bm{pks}$ vector.
In a similar manner, the interval between $p_{i}$ and $M_{1}$ is again divided to obtain the pAC as:
\begin{equation}
M_{2} = \dfrac{p_{i} + M_{1}}{2} = \dfrac{3p_{i} + p_{i+1}}{4}\label{equation}
\end{equation}
	
Now, the SCG signal is segmented from $M_{2}$ to $M_{1}$ for each of the cycles. Subsequently, the DC-offset is removed and amplitude normalization is performed. Subsequently, moving average filtering is applied to reduce irregularities. To detect all pAC points, the trends in the segments are removed by applying a Butterworth high pass IIR filter with an empirical cut-off frequency of $10$ Hz. Finally, the pAC peaks are detected as maximas of the processed SCG segments. After annotating AO and pAC peaks, HR and LVET$'$ interval for each of the beat are calculated. Using these LVET$'$ and HR values for each individual subject, the system is calibrated with the original SBPs and DBPs using a linear regression model. Hence, model coefficients are obtained for both SBPs and DBPs, separately. The coefficients obtained for a particular subject are finally used  to estimate SBP and DBP values. 
\begin{figure}
\centering
\includegraphics[width=3.3in]{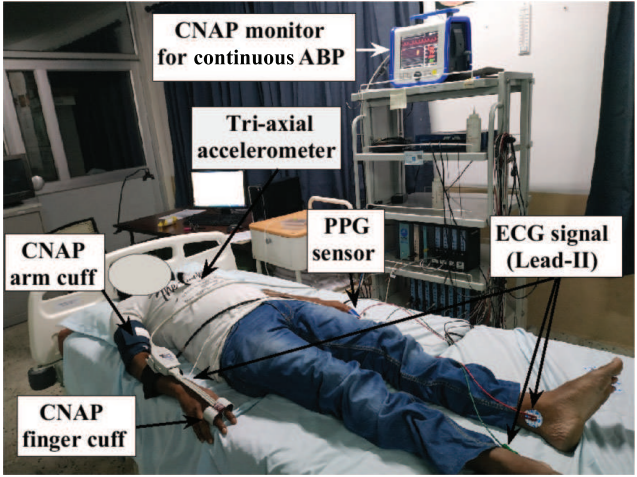}
\caption{Experimental setup for signal acquisition.}
\label{experiment_set}
\end{figure}

\section{Experiments and Results}
 Fig. \ref{experiment_set} shows the experimental setup for simultaneous recording of SCG, ECG, PPG and ABP signals.
Data was recorded for $6$ minutes from 10 healthy subjects in supine position. For all the subjects, recordings were done for ECG in Lead-II configuration, z-axis signal of SCG using tri-axial accelerometer, PPG at fingertip, and ABP using finger-cuff provided by CNAP\textsuperscript{\textregistered} Monitor 500. All signals are synchronously obtained using  MP150 DAQ system (BIOPAC Systems, Inc.) at a sampling rate of 1 kHz. Data acquisition is performed at Electro Medical and Speech Technology (EMST) Laboratory, IIT Guwahati with prior consent of the subjects after clearance from institutional ethics review board. For each subject, $70\%$  of the data is used for calibration, while the remaining is used for the testing purpose. The ABP signal was used for calibration of the SBP and DBP of an individual. It is used as a benchmark to test and validate the proposed system. 
\begin{table}
\caption{BP and HR ranges in the database}
\centering{
\label{Stat_BP_HR}
\begin{tabular}{ccccc}
\toprule
\multicolumn{1}{l}{} & \textbf{Min} & \textbf{Max} & \textbf{STD} & \textbf{Mean} \\ \midrule
\textbf{DBP (mmHg)}           & 55.16        & 97.33        & 6.749        & 74.180        \\ 
\textbf{SBP (mmHg)}           & 78.16        & 161.01       & 13.66        & 122.84        \\ 
\textbf{HR (bpm)}            & 49.2         & 141.17       & 11.8         & 81            \\ \bottomrule
\end{tabular}}
\end{table}
Table \ref{Stat_BP_HR} shows statistical distribution in terms of minimum (Min), maximum (Max), standard deviation (STD) and mean values of the BP and HR for in-house recorded database. The Min and Max SBP values are $78.16$ mmHg and $161.01$ mmHg, whereas for DBP, Min and Max values are $55.15$ mmHg and $97.33$ mmHg, respectively. Min recorded HR is $49.2$ bpm, while max recorded HR in the in-house data  is $141.17$ bpm. For each individual subject, $70\%$ of the data is used for calibration and the algorithm is evaluated on the remaining data. Mean error (ME), mean absolute error (MAE) and standard deviation (STD) are used to compare the performance with the original values. 
BP is estimated using the proposed LVET$'$-based approach. The estimated SBP and DBP values are compared with ABP measurements. The performance of the proposed approach is evaluated in terms of ME, MAE and STD measures, and the  results are shown in Table \ref{LVET_BP_ABP}.    
\begin{table}
\caption{Comparison of estimated BPs using our proposed LVET$'$-based method with reference ABP measurements}
\centering{
\label{LVET_BP_ABP}
\begin{tabular}{ccclccl}
\toprule
\multirow{2}{*}{\textbf{Subjects}} & \multicolumn{3}{c|}{\textbf{DBP (mmHg)}} & \multicolumn{3}{c}{\textbf{SBP (mmHg)}}                          \\ 
\cmidrule(r){2-4}                                                                                                                                                                                                                    \cmidrule(l){5-7}                                                                                                                                                                                                                  
& \textbf{$\textbf{ME}$} & \textbf{$\textbf{MAE}$} & \multicolumn{1}{c|}{\textbf{$\textbf{STD}$}} & \textbf{$\textbf{ME}$} & \textbf{$\textbf{MAE}$} & \multicolumn{1}{c}{\textbf{$\textbf{STD}$}} \\  \midrule
1                                  & -1.66        & 1.98          & 2.15                               & 1.09         & 2.69          & 3.28                               \\ 
2                                  & -2.93        & 2.93          & 1.96                               & -2.81        & 3.07          & 2.25                               \\ 
3                                  & -1.50        & 2.99          & 3.33                               & 0.57         & 3.03          & 3.86                               \\ 
4                                  & 1.78         & 2.54          & 2.75                               & $\approx 0.00$         & 2.14          & 2.65                               \\ 
5                                  & -0.07        & 1.88          & 2.11                               & 0.44         & 1.04          & 1.41                               \\ 
6                                  & -0.70        & 1.95          & 2.06                               & -0.62        & 2.50          & 3.07                               \\ 
7                                  & -1.22        & 2.31          & 2.60                               & 1.95         & 4.73          & 5.42                               \\ 
8                                  & -2.03        & 2.66          & 2.19                               & -4.99        & 4.99          & 2.91                               \\ 
9                                  & -2.17        & 3.51          & 3.73                               & -0.39        & 1.72          & 2.11                               \\ 
10                                 & -2.43        & 3.22          & 2.92                               & 2.88         & 5.86          & 6.39                               \\
\midrule
\textbf{Average}                                  & -1.29        & 2.6 & 2.6                               & -0.19         & 3.2          & 3.3\\
\bottomrule
\end{tabular}}
\end{table}
\begin{table*}[t!]
\caption{Statistical comparison of the proposed LVET$'$-based method with the PTT and PTT1 based methods for DBP and SBP measurements}
\resizebox{18cm}{!}{%
\centering{
\label{Developed-Existed}
\begin{tabular}{cccllllllccllllll}
\toprule
\multirow{2}{*}{\textbf{Subjects}} & \multicolumn{8}{c}{\textbf{DBP (mmHg)}}                                                                                                                                                                                                               & \multicolumn{8}{c}{\textbf{SBP (mmHg)}}      
 \\ \cmidrule(r){2-9} 
\cmidrule(l){10-17}
                                   & $\textbf{ME}_\textbf{PTT}$ & $\textbf{ME}_\textbf{PTT1}$ & \multicolumn{1}{c}{$\textbf{MAE}_\textbf{PTT}$} & \multicolumn{1}{c}{$\textbf{MAE}_\textbf{PTT1}$} & \multicolumn{1}{c}{$\textbf{STD}_\textbf{PTT}$} & \multicolumn{1}{c}{$\textbf{STD}_\textbf{PTT1}$} & \multicolumn{1}{c}{\textbf{r1}} & \multicolumn{1}{c}{\textbf{r2}} & $\textbf{ME}_\textbf{PTT}$ & $\textbf{ME}_\textbf{PTT1}$ & \multicolumn{1}{c}{$\textbf{MAE}_\textbf{PTT}$} & \multicolumn{1}{c}{$\textbf{MAE}_\textbf{PTT1}$} & \multicolumn{1}{c}{$\textbf{STD}_\textbf{PTT}$} & \multicolumn{1}{c}{$\textbf{STD}_\textbf{PTT1}$} & \multicolumn{1}{c}{\textbf{r1}} & \multicolumn{1}{c}{\textbf{r2}} \\ \midrule
1                                  & -0.26        & -0.02        & 0.35                               & 0.13                               & 0.31                               & 0.16                               & 0.44                             & 0.69                             & -0.42        & -0.06        & 0.59                               & 0.34                               & 0.53                               & 0.42                               & 0.01                             & 0.76                             \\ 
2                                  & 0.02         & 0.02         & 0.07                               & 0.11                               & 0.09                               & 0.16                               & 0.94                             & 0.83                             & -0.04        & 0.02         & 0.27                               & 0.24                               & 0.36                               & 0.39                               & 0.58                             & 0.58                             \\ 
3                                  & 0.18         & -0.01        & 0.57                               & 0.12                               & 0.77                               & 0.19                               & 0.66                             & 0.97                             & 0.76         & -0.01        & 1.81                               & 0.07                               & 2.56                               & 0.12                               & 0.49                             & 1.00                             \\ 
4                                  & -0.87        & -0.16        & 1.17                               & 0.28                               & 1.40                               & 0.62                               & 0.10                             & 0.46                             & -0.32        & -0.09        & 0.45                               & 0.22                               & 0.55                               & 0.50                               & 0.78                             & 0.78                             \\ 
5                                  & 1.04         & 0.13         & 1.36                               & 0.51                               & 1.16                               & 0.71                               & 0.05                             & 0.27                             & 0.22         & 0.02         & 0.29                               & 0.07                               & 0.25                               & 0.09                               & 0.31                             & 0.67                             \\ 
6                                  & 0.02         & -0.02        & 0.03                               & 0.14                               & 0.02                               & 0.18                               & 0.99                             & 0.56                             & 0.66         & -0.04        & 0.69                               & 0.32                               & 0.45                               & 0.42                               & 0.37                             & 0.31                             \\ 
7                                  & 0.05         & -0.20        & 0.10                               & 0.21                               & 0.11                               & 0.11                               & 0.71                             & 0.68                             & 0.10         & 0.61         & 0.13                               & 0.61                               & 0.12                               & 0.10                               & 0.55                             & 0.66                             \\ 
8                                  & 0.24         & 0.00         & 0.71                               & 0.08                               & 0.80                               & 0.12                               & 0.11                             & 0.56                             & 0.23         & -0.75        & 0.65                               & 0.75                               & 0.73                               & 0.10                               & 0.04                             & -0.09                            \\ 
9                                  & -0.02        & -0.29        & 0.16                               & 0.30                               & 0.28                               & 0.27                               & 0.70                             & 0.72                             & 0.02         & 3.37         & 0.08                               & 3.37                               & 0.10                               & 0.38                               & 0.74                             & -0.33                            \\ 
10                                 & 0.04         & -0.93        & 0.69                               & 1.02                               & 1.22                               & 1.00                               & 0.80                             & 0.77                             & -0.02        & 4.08         & 0.26                               & 4.58                               & 0.58                               & 3.24                               & 0.99                             & 0.82                             \\ \midrule
\textbf{Average}                                  & 0.04       & -0.15 & 0.52                               & 0.29                               & 0.62                               & 0.35                               & 0.55                             & 0.65                             & 0.12        & 0.72        & 0.52                               & 1.06                               & 0.62                               & 0.58                               & 0.49                             & 0.52                             \\ 
\bottomrule
\end{tabular}}}
\end{table*}
For the DBP measurement, the minimum ME and MAE values are $-0.07$ and $1.88$, respectively for subject 5 with a STD value of $2.11$. Similarly, for the SBP measurement, minimum ME is closed to zero for subject 4 with a STD value $2.65$.  The minimum MAE value is $1.04$ for subject 5 with a STD value $1.41$. For all the subjects, the average ME with STD of the estimated DBP and SBP are $-1.29\pm2.6$ mmHg and $-0.19\pm3.3$ mmHg, respectively, whereas the average MAE are, $2.6$ and $3.2$ mmHg, respectively. The obtained results suggest that the accuracy of BP estimations using the proposed system satisfies the requirements of the IEEE standard $5\pm8$ mmHg \cite{b7, b8}.

The estimated BP values are also compared with the BP values computed through earlier PTT- and PTT1-based approaches. In PTT-based approach, the duration of PTT is calculated using the combination of `ECG-PPG' \cite{b2,b8,b9,b10,b11,b12}, whereas in the PTT1-based approach, `SCG-PPG' signal pair is used \cite{b5,b14,b15}. Table \ref{Developed-Existed} shows the comparison of the proposed method with the existing methods.
The average ME and MAE between the proposed LVET$'$-based and PTT-based methods are found as $0.04\pm0.62$ mmHg and $0.52$ mmHg for DBP, and $0.12\pm0.62$ mmHg and $0.52$ mmHg for SBP measurement, respectively. While, for the PTT1-based method, these metrics are observed as $-0.15\pm0.35$ mmHg and $0.29$ mmHg for DBP measurement. While for SBP, they are found as $0.72\pm0.58$ mmHg and $0.58$ mmHg, respectively.
Additionally, the average correlation of the BPs estimated from the proposed method with that of the earlier methods is also computed. For the PTT-based method, the average correlation coefficients, denoted by $r1$, are obtained as 0.6 and 0.5 for the DBP- and SBP-estimation, respectively. Similarly, for PTT1-based method, the correlation coefficient ($r2$) values are observed as 0.7 and 0.5 for the DBP and the SBP, respectively. The good correlation exhibited by the proposed method  clearly shows that it is comparable with the existing methods. Therefore, a single sensor based proposed method may replace the use of multi-modal system for the BP estimation.

\begin{figure}
\centerline{\includegraphics[width=3.6in]{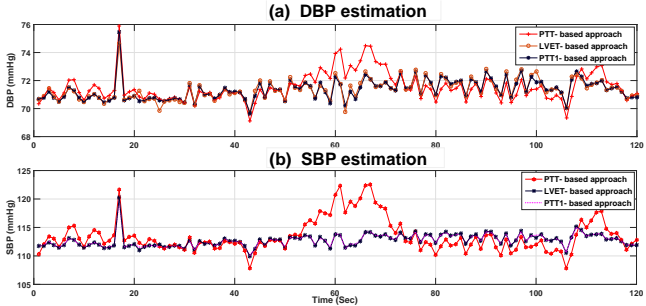}}
\caption{Comparison of the proposed LVET$'$ based method with PTT and PTT1 based methods for estimated DBP and SBP values.}
\label{DBP-SBP}
\vspace{-0.3cm}
\end{figure}
Further, to show the performance and its comparison with the existing methods, the BP traces estimated from a single subject (\#3) are taken for the study. The comparison is shown in terms of Bland-Altman and regression plots. For subject 3, the traces of beat-to-beat DBP and SBP values estimated by the proposed and the existing methods are shown in Fig.~\ref{DBP-SBP}. It is observed that the proposed method and the existing state-of-the-art methods produce similar DBP and SBP measurements. Hence, it shows the ability of our method to be used for continuous long term BP estimation with a single cardiac sensor itself.

\begin{figure*}[t!]
\centering
\includegraphics[width=3.4in]{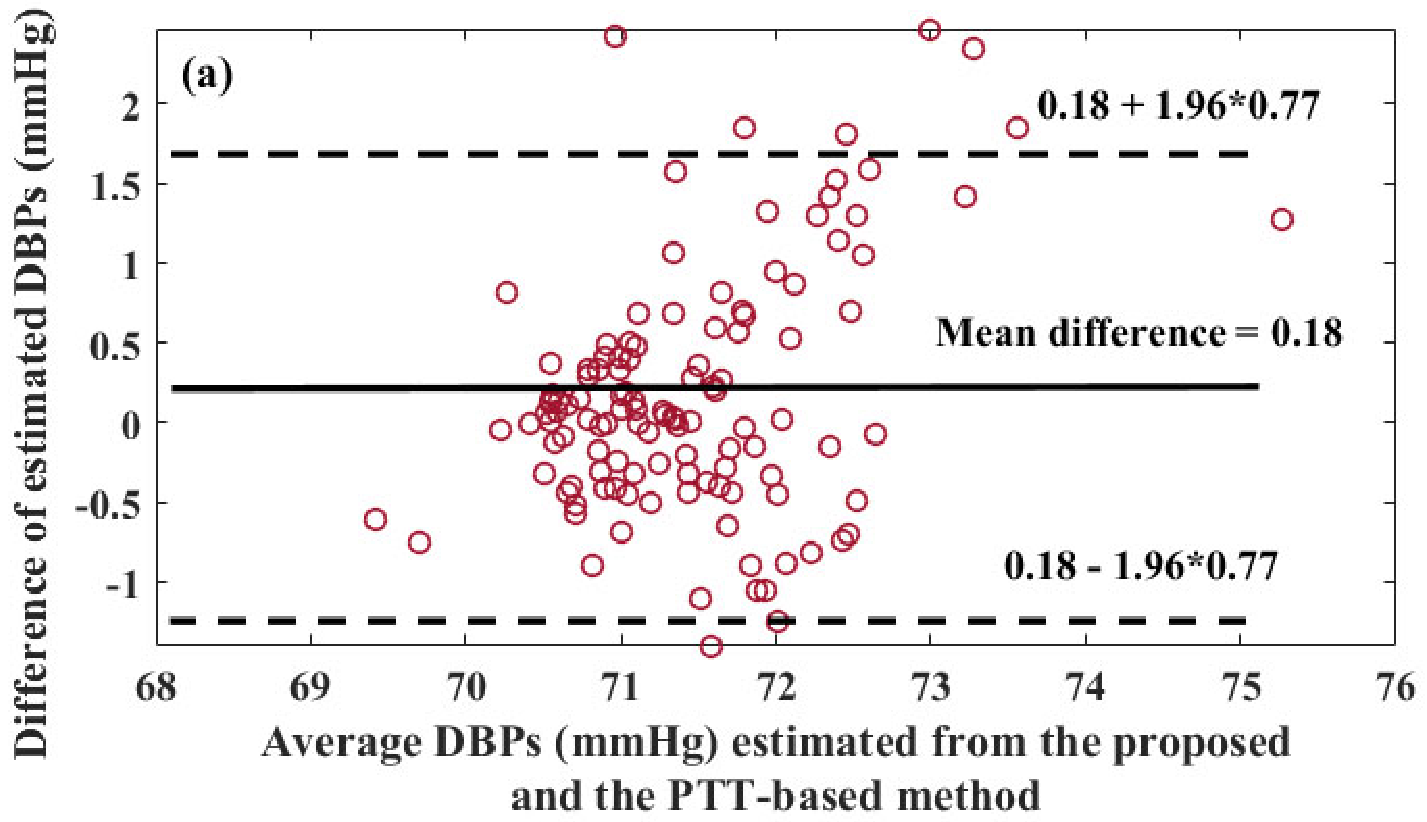}
~~~~\includegraphics[width=3.4in]{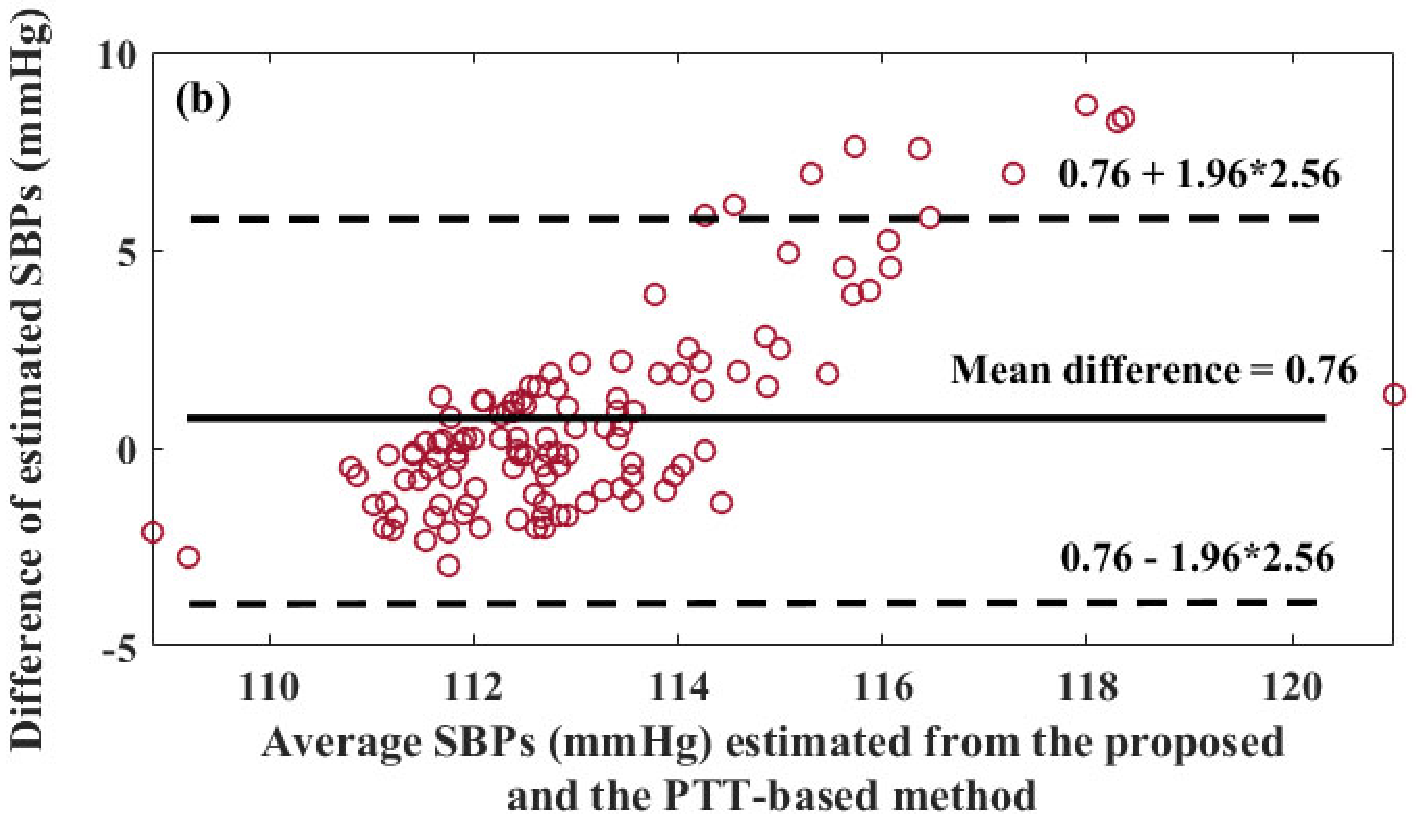}

\includegraphics[width=3.4in]{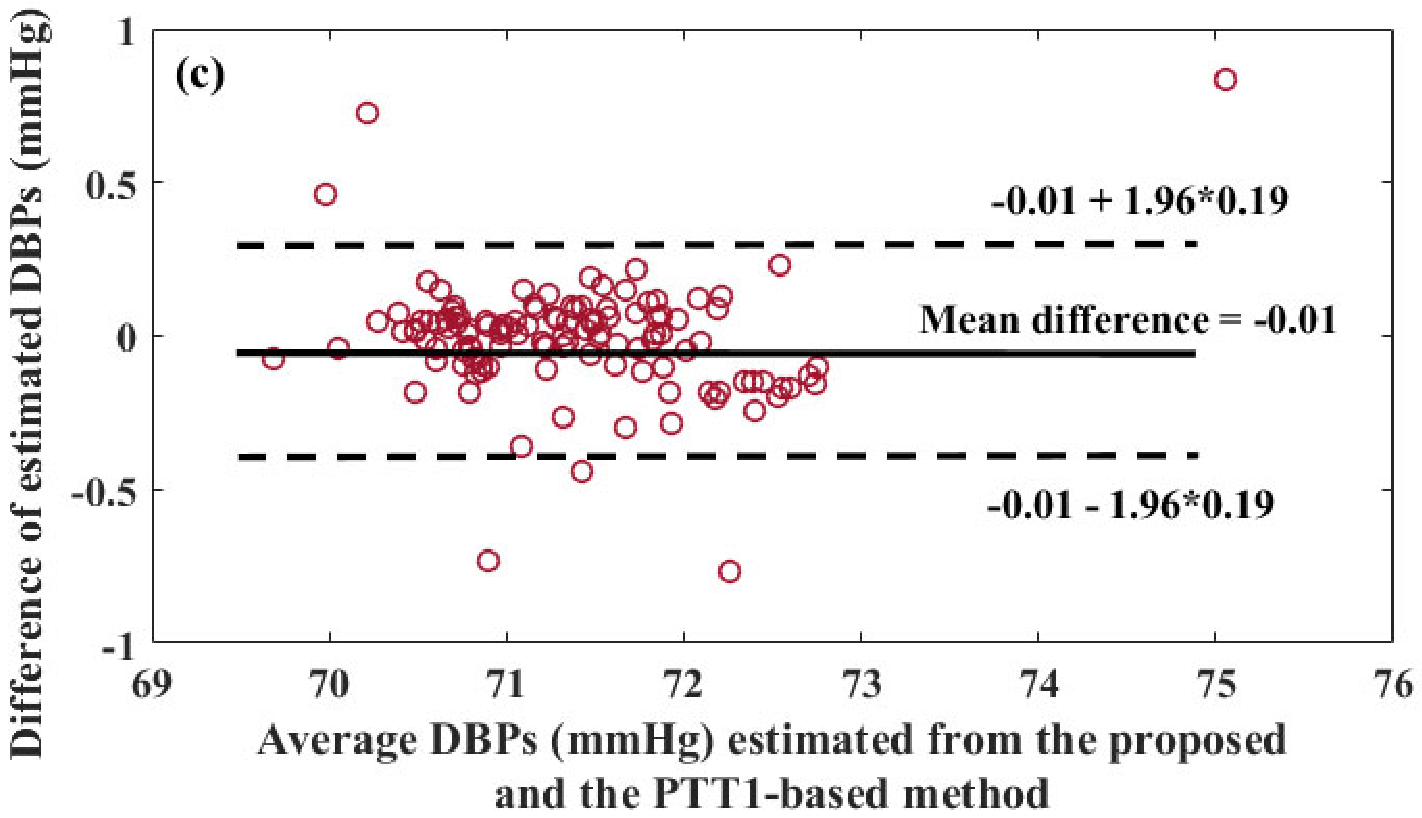}
~~~~\includegraphics[width=3.4in]{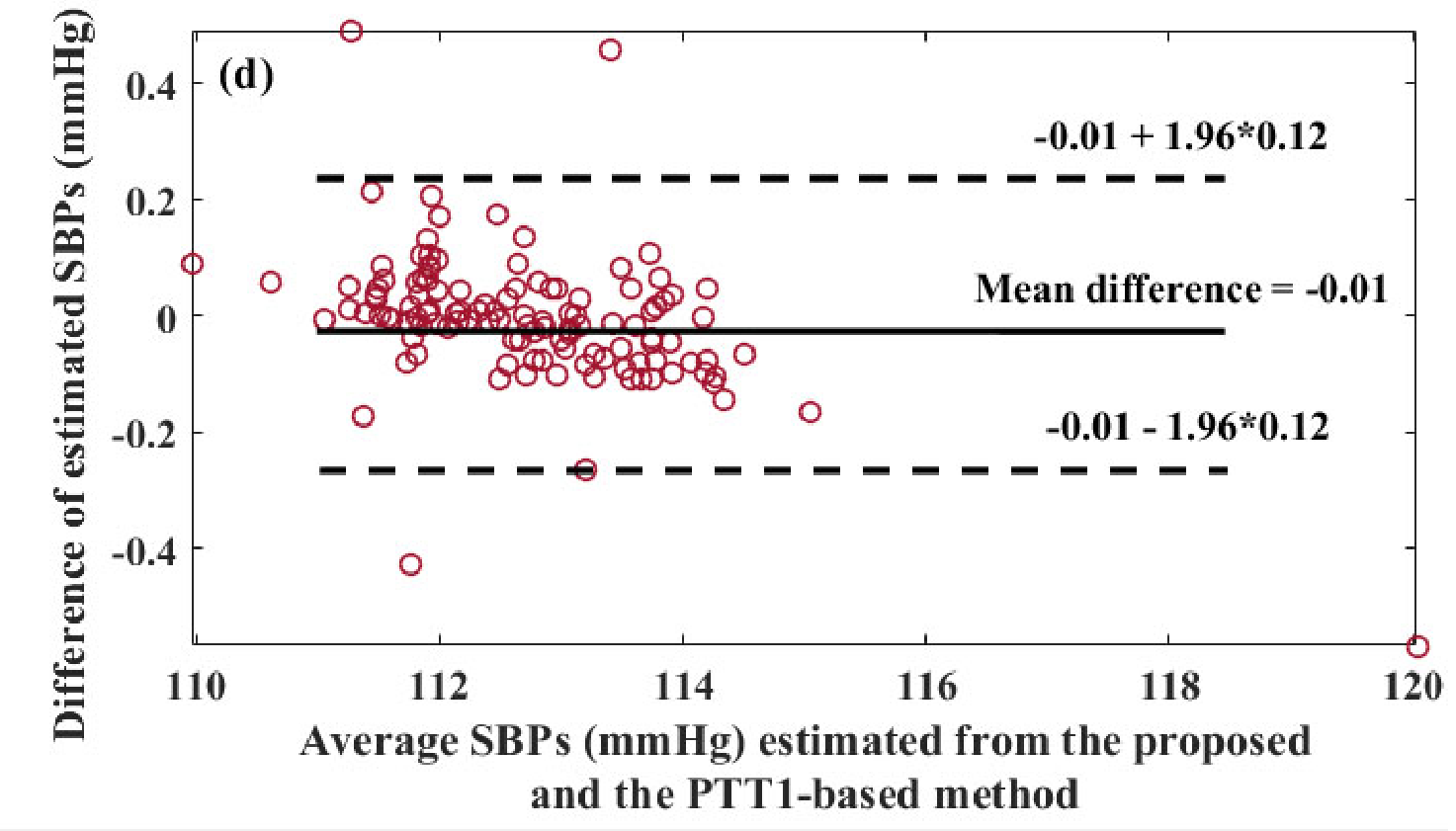}
\caption{Bland-Altman plot to compare the performance of the proposed method with PTT- and PTT1-based methods. Panels (a) and (b) represent comparison with PTT-based method for DBPs and SBPs, respectively, and panels (c) and (d) represent comparison with PTT1-based method for DBPs and SBPs, respectively.}
\label{BAplot}
\end{figure*}


\begin{figure*}[t!]
\centering
\includegraphics[width=3.3in]{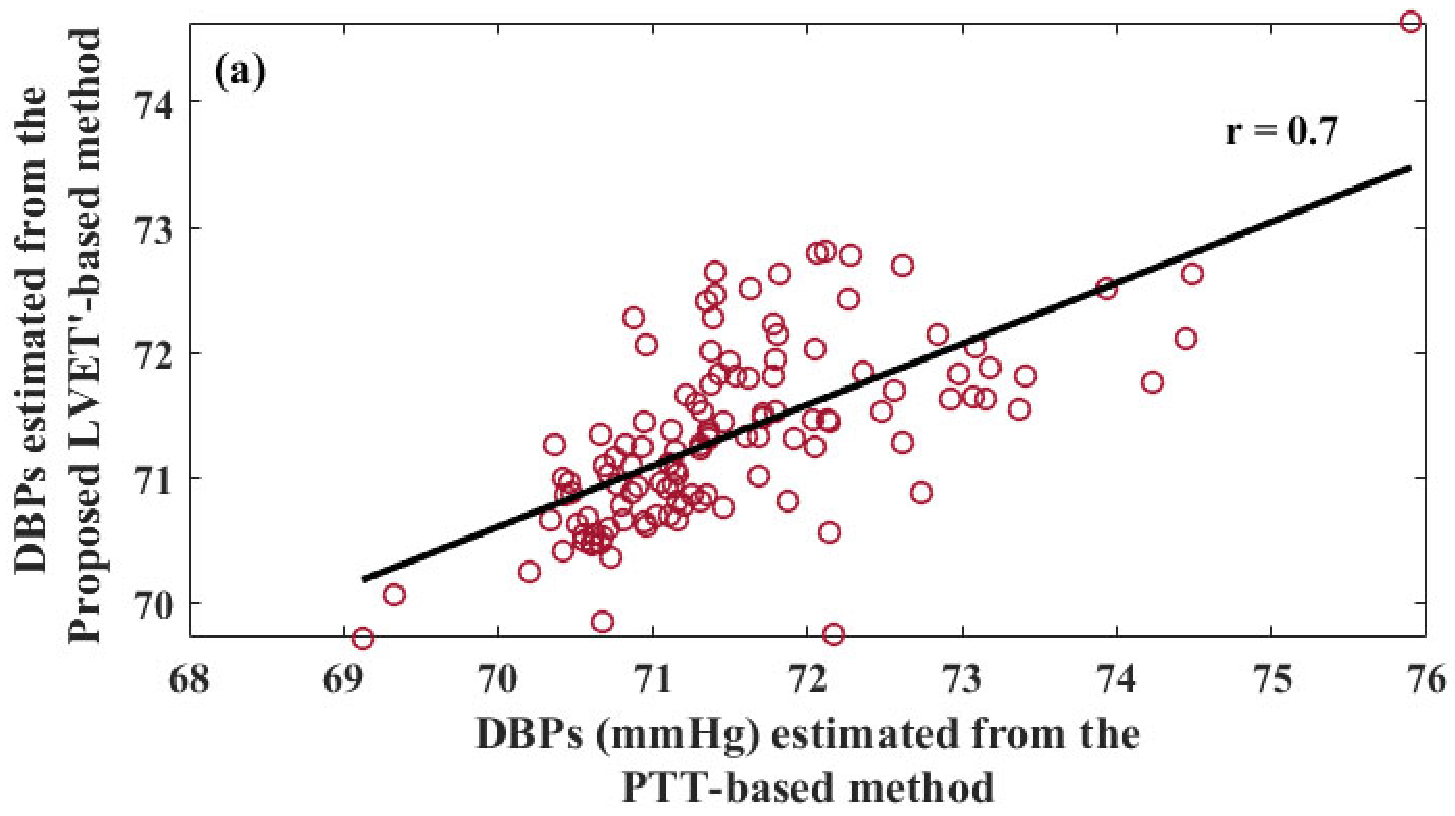}
~~~~\includegraphics[width=3.3in]{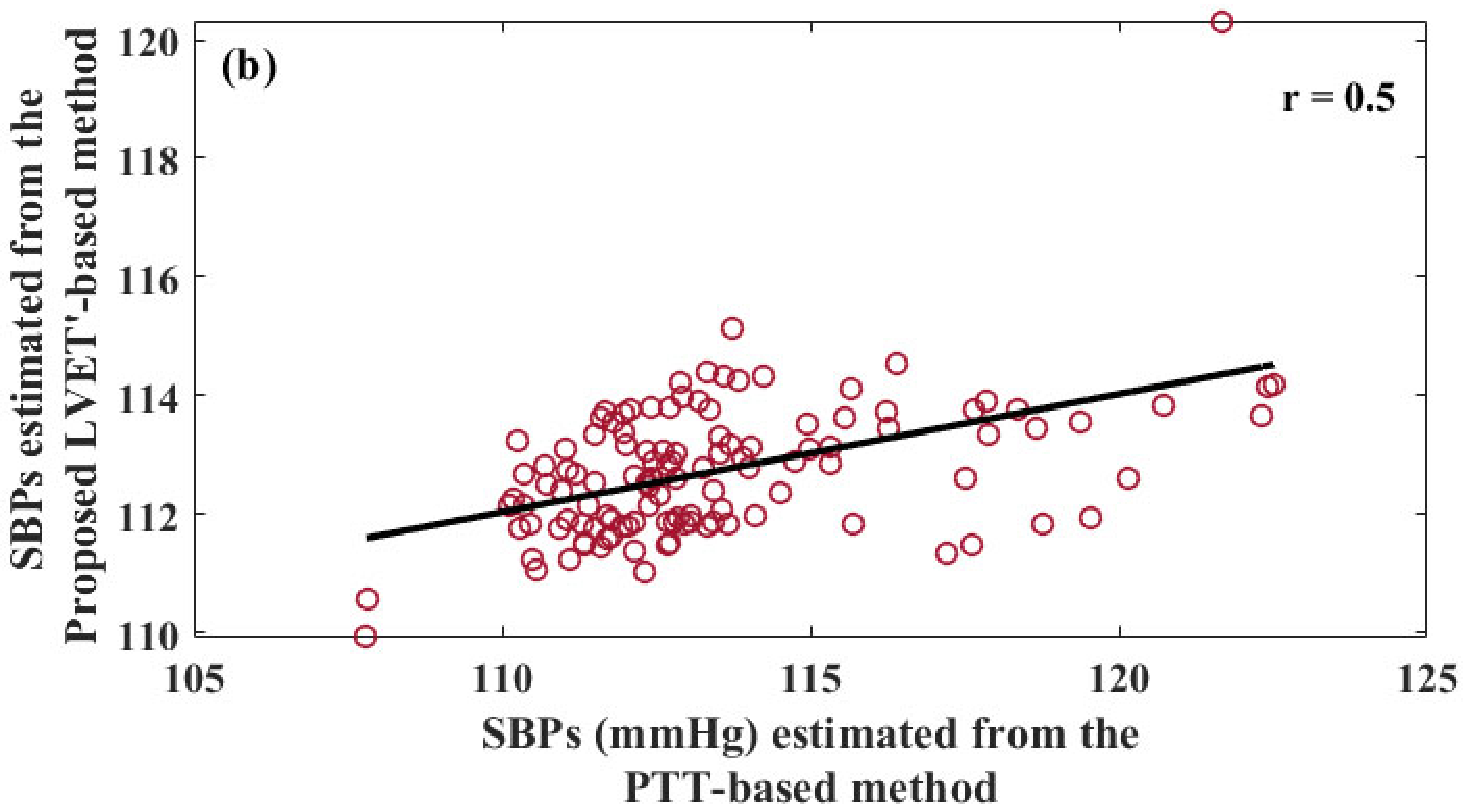}

\includegraphics[width=3.3in]{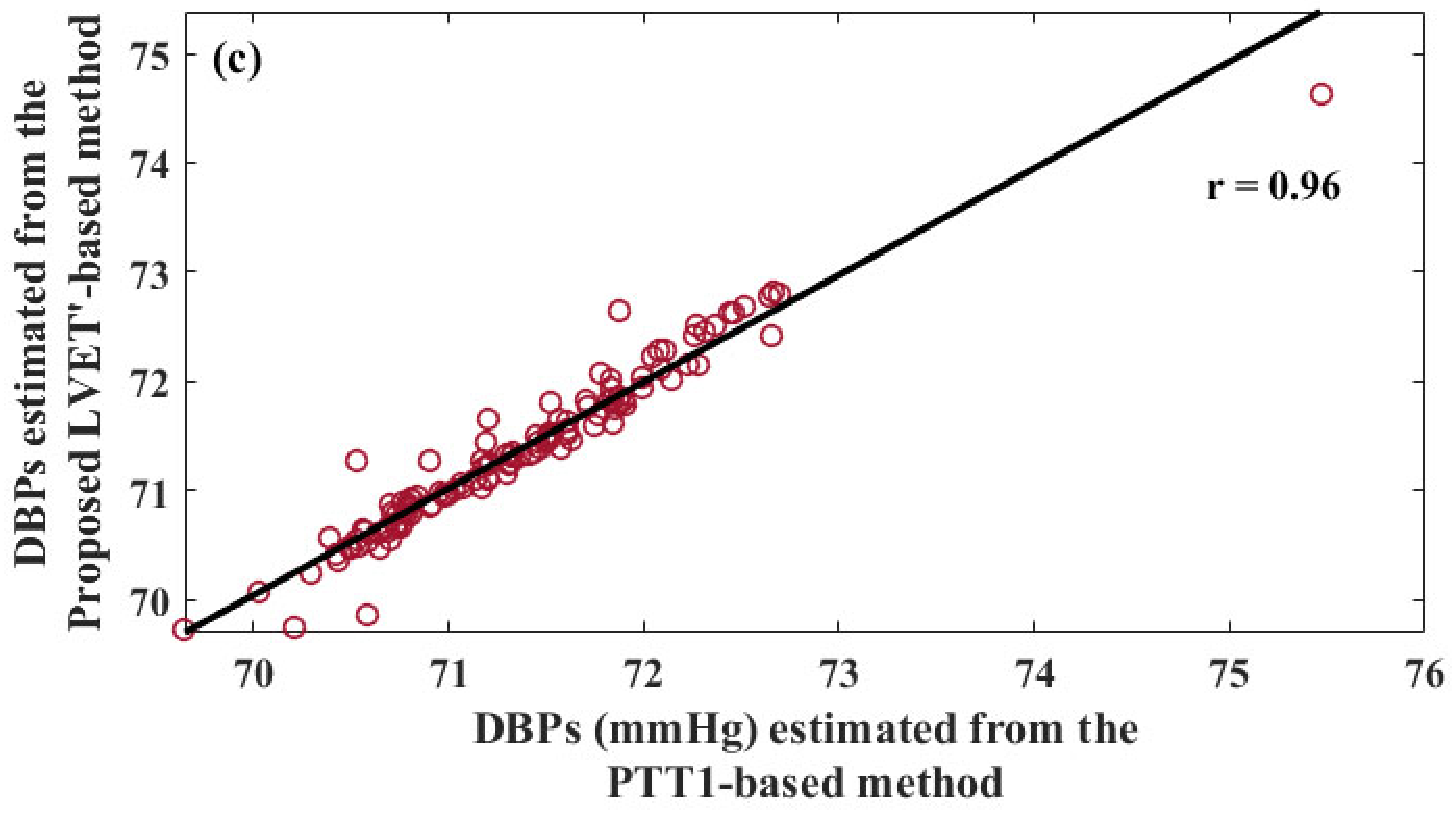}
~~~~\includegraphics[width=3.3in]{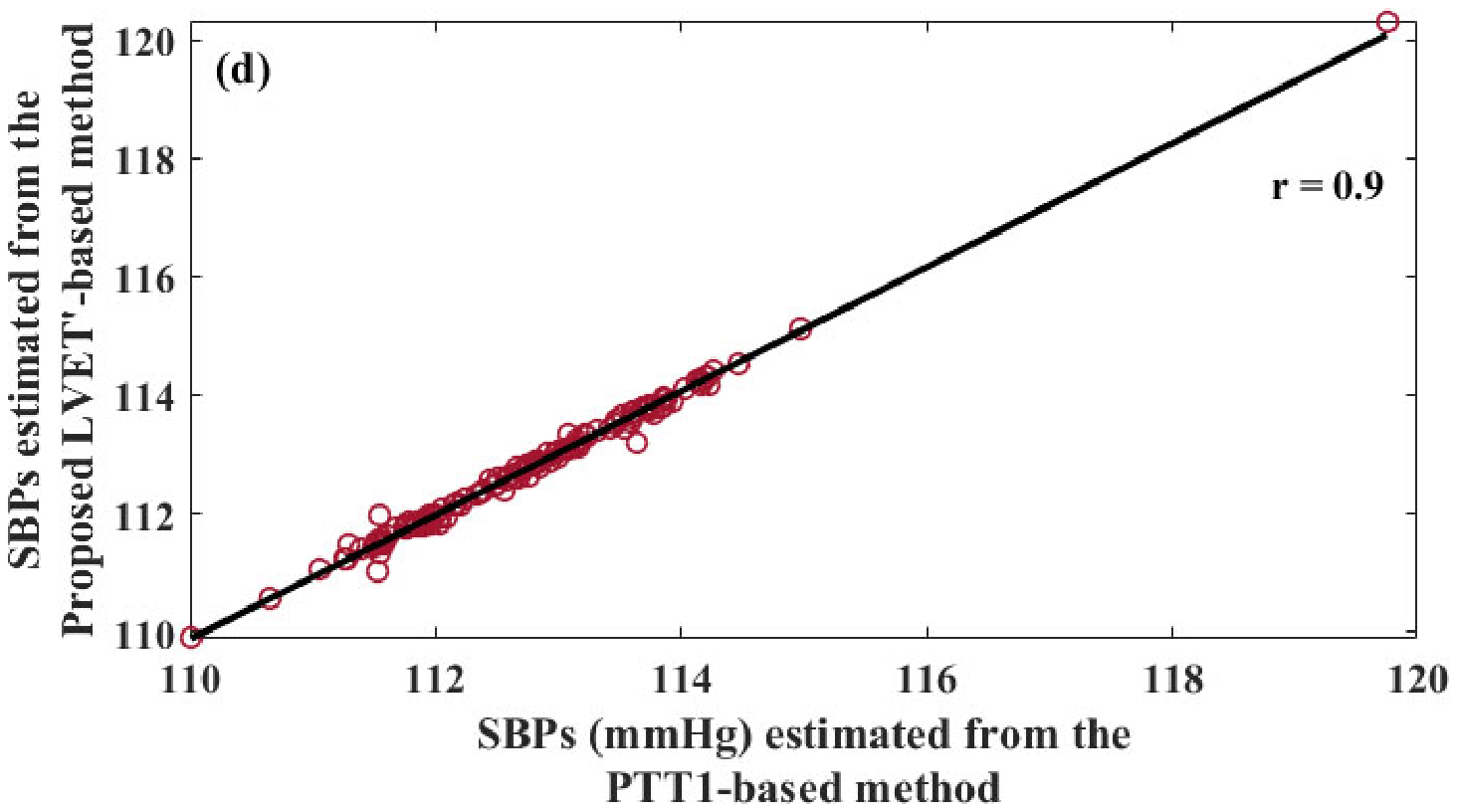}
\caption{Regression plots of estimated BP measurements for a single subject. (a) and (c) show the regression of DBPs estimated by the proposed method with that of the PTT- and PTT1-based methods, respectively, (b) and (d) show the regression of SBPs estimated by the proposed method with that of the PTT- and PTT1-based methods, respectively.}
\label{plot_regression}
\end{figure*}


Fig. \ref{BAplot} shows the Bland-Altman plots for performance comparison in terms of measurement-difference in BP estimation. The $x$-axis represents the average of the BP values obtained by the proposed and the existing method, whereas the difference between both the BP values is shown in the $y$-axis. It is desirable to have a low value of bias or mean difference. This result indicates that  the proposed method can give similar measurement values as that of the method used for comparison of performance. As shown by the bold line in Fig. \ref{BAplot}, the mean differences of the estimated DBPs and SBPs between the proposed and the PTT-based methods are $0.18\pm0.77$ mmHg and $0.76\pm2.56$ mmHg, respectively. Whereas, they are found to be $-0.01\pm 0.19$ mmHg and $-0.01\pm 0.12$ mmHg, respectively for the PTT1-based method. It can be observed that the majority of the points lie within the limits of agreement, \textit{i.e.,} $m \pm1.96 s$, where $m$ and $s$ denote the mean  and standard deviation of the BP difference, respectively. The limits of agreement are shown  by dashed lines in the figure. This shows the suitability of our method as a substitute of the existing methods.

In addition, the performance comparison is also shown with the help of regression plots in Fig. \ref{plot_regression}. The corresponding correlation coefficients between the proposed LVET$'$-based and the PTT based methods are approximately 0.7 and 0.5  for the DBP and SBP estimation, respectively, as shown in Fig. \ref{plot_regression}(a) and (b). Whereas, in case of the PTT1-based method, it is found 0.9 for both the DBPs and SBPs as shown in Fig. \ref{plot_regression}(c) and (d). 

\section{Conclusion}
In this work, a method is proposed for noninvasive and cuffless long-term BP measurement using an accelerometric SCG signal. 
Initially, the AO and pAC peaks are detected for the estimation of LVET$'$ and HR parameters. With these parameters, the BP modeling is performed for the SBPs and the DBPs using individual calibration. The existing methods require a PPG signal along with either ECG or SCG signals to perform this task.  However, the use of bio-electrodes for the ECG and the photo-detector for PPG acquisition has been avoided in the proposed method by using only SCG data.
This makes it more flexible and offers mobility to the user. Also, the light-weight of the sensing device and its simple operation make it versatile to be operated independently without any help of an expert.
With in-house recordings, quite satisfactory correlation results between the proposed method and the existing methods are obtained.
The performance results clearly show that our method is able to achieve the performance up to the level obtained by the state-of-the-art methods only with a single cardiac sensor.  
In this research, the proposed method is tested and validated on healthy individuals under a normal condition, and hence, the performance could be evaluated on different physiological and pathological conditions in the future works. However, the quantitative results show the potential of the proposed method to replace the existing methods for continuous long-term BP monitoring.

\end{document}